\documentclass[11pt,paper]{JHEP3}
\usepackage{amsmath}
\usepackage{bm}
\usepackage{times}
\usepackage{epsfig}

\title{Dark Energy from a wet dark fluid}

\author{R.~Holman and Siddartha Naidu \\ 
  Department of Physics, Carnegie Mellon University, Pittsburgh PA, USA
\email{rh4a@andrew.cmu.edu,svn@andrew.cmu.edu}}

\abstract{ We propose a new equation of state for the Dark Energy
component of the Universe. It is modeled on the equation of state
$p=w(\rho-\rho_{*})$ which can describe a liquid, for example
water. We show that its energy density naturally decomposes into a
component that behaves as a cosmological constant and one whose energy
density scales as $a^{-3(1+w)}$, and fit the parameters specifying the
equation of state to the new SNIa data, as well as WMAP and 2dF
data. We consider both the case where the dark fluid is smooth ({\it i.e} only the CDM component clusters gravitationally) as well as the case where the dark fluid also clusters. We find that for both cases, reasonable values of the parameters can be found
that give our model the same $\chi^2$ as that of $\Lambda$CDM. Furthermore, in the case where our dark fluid clusters, allowing a blue tilt to the power spectrum allows us to fit all the requisite data with the dark fluid being the dominant component of the dark matter of the Universe. A
remarkable feature of the model is that we can do all this with a microphysical
$w>0$. We also display a field theoretic model that yields this equation
of state.}

\keywords{cos,ctg}
\begin{document}

The nature of the dark energy component of the Universe
\cite{sst,scp,review} remains one of the deepest mysteries of
cosmology. There is certainly no lack of canditates: a remnant
cosmological constant, quintessence\cite{quint}, k-essence\cite{kess},
phantom energy\cite{phantom}. Modifications of the Friedmann equation
such as Cardassian expansion\cite{card} as well as what might be
derived from brane cosmology\cite{brane} have also been used to try to
explain the acceleration of the Universe.

In this work, we offer a new candidate for the dark energy: Wet Dark
Fluid (WDF). This model is in the spirit of the generalized Chaplygin gas (GCG)
\cite{GCG}, where a physically motivated equation of state is offered with
properties relevant for the dark energy problem. Here the motivation
stems from an empirical equation of state proposed by Tait
\cite{tait,hayward} in 1888 to treat water and aqueous solutions.
The equation of state for WDF is very simple,
\begin{equation}
\label{cdfeos}
p_{\rm WDF} = w(\rho_{\rm WDF}-\rho_{*}),
\end{equation}
and is motivated by the fact that this is a good approximation for
many fluids, including water, in which the internal attraction of the
molecules makes negative pressures possible. One of the virtues of this model is that the square of the sound
speed, $c_s^2$, which depends on $\partial p/\partial \rho$, can be positive (as opposed to the case of phantom energy, say), while still giving rise to cosmic acceleration in the current epoch. 

In real fluids negative pressures eventually lead to a breakdown of
Eq.~(\ref{cdfeos}) due to cavitation \cite{cav}, but for now we simply
treat Eq.~(\ref{cdfeos}) as a phemenological equation\cite{chiba}.  We will show that
this model can be made consistent with the most recent SNIa
data\cite{Riess}, the WMAP results\cite{WMAP} as well as the
constraints coming from measurements of the matter power
spectrum\cite{2df}.  The parameters $w$ and $\rho_{*}$ are taken to be
positive and we restrict ourselves to $0<w<1$. Note that if $c_s$
denotes the adiabatic sound speed in WDF, then $w=c_s^2$.

To find how the WDF energy density scales with the scale factor
$a$, we use the energy conservation equation together with the
equation of state in Eq.~(\ref{cdfeos}):
\begin{equation}
\label{cdfscale}
\begin{split}
 &\dot{\rho}_{\rm WDF}+ 3 H (p_{\rm WDF} +\rho_{\rm WDF})=0 \\
\Rightarrow & \rho_{\rm WDF}=\frac{w}{1+w} \rho_{*}+ D(\frac{a_0}{a})^{3(1+w)},
\end{split}
\end{equation}
where $D$ is a constant of integration and $a_0$ is the scale factor
today; we will set $a_0=1$ from now on.

WDF naturally includes two components: a piece that behaves as a
cosmological constant as well as a piece that redshifts as a standard
fluid with an equation of state $p=w \rho$. We can show that {\em if}
we take $D>0$, this fluid will never violate the strong energy condition
$p+\rho\geq 0$:
\begin{equation}
\label{wec}
\begin{split}
p_{\rm WDF} +\rho_{\rm WDF}&=(1+w)\rho_{\rm WDF}-w\rho_{*}\\
                           &=D (1+w)(\frac{a_0}{a})^{3(1+w)}\geq 0.
\end{split}
\end{equation}

It is tempting to try to use the second component as the dark
matter, thus unifying the two dark components. There is the potential problem that since the sound
speed of the second component is non-zero, this would give rise to a pressure
gradient term in the equation for linear fluctuations. At least in the case of the Chaplygin gas, this has been shown to slow down the
growth of fluctuations to a level that would be inconsistent with
measurements of the power spectrum\cite{sandviketal} as well as with the CMB
fluctuations (although see \cite{hu,bertolami} for possible ways out
of this predicament). We will start off by assuming that WDF does {\em not}
cluster gravitationally and that the formation of structure is driven,
as in the standard $\Lambda$CDM model, by the clustering of a cold
dark matter component (this was done in ref.\cite{gaztanaga} for the
GCG model). Later on in the paper, we will examine whether and to what extent fluctuations in the matter power spectrum (MPS) are suppressed if we allow fluctuations in the WDF fluid and whether adjusting the tilt of the power spectrum could reduce such a suppresion. 

The Friedmann and acceleration equations for the WDF/CDM system in a
spatially {\em flat} FRW universe are:
\begin{equation}
\label{friedaccel}
\begin{split}
H^2 = H_0^2 & \bigl[  \Omega_{CDM}\ a^{-3} 
                    + \Omega_{\rm WDM}\ a^{-3(1+w)}
                    + \Omega_{\rm WDE} \bigr] \\
\frac{\ddot{a} }{a} = -\frac{H_0^2}{2} &
  \bigl[ \Omega_{\rm CDM}\  a^{-3} -  2 \left(\Omega_{\rm WDE}+\Omega_{\rm WDM}\right) \\
     & \quad +3(1+w) \Omega_{\rm WDM}\  a^{-3(1+w)} \bigr],
\end{split}
\end{equation}
where WDE, WDM stand for wet dark energy/matter respectively and
\begin{equation}
\label{omegas}
\Omega_{\rm WDE}  =  \frac{8\pi G_N}{3 H_0^2} \frac{w}{1+w} \rho_{*} ,\ \ \Omega_{\rm WDM}  =  \frac{8\pi G_N}{3 H_0^2} D,
\end{equation}
with $\Omega_{\rm WDE}+\Omega_{\rm WDM}+\Omega_{\rm CDM}  =  1$.
We can also rewrite the equation of state in the more traditional form $p_{\rm WDF} = w_{\rm eff}(z) \rho_{\rm WDF}$ where
\begin{equation}
\label{weff }
\begin{split}
w_{\rm eff}(z) = & \frac{w\ \Omega_{\rm WDM}\ (1+z)^{3(1+w)}-\Omega_{\rm WDE}}{\Omega_{\rm WDM} \ (1+z)^{3(1+w)}+\Omega_{\rm WDE}}.
\end{split}
\end{equation}
We see that $w_{\rm eff}$ interpolates between $w$ at early times and
$-1$ at late times. Note that this is of a similar form to that used
in \cite{hannestad}, except that some of the parameters in their
$w_{\rm eff}$ that they take to be positive are in fact negative in
our version.

\EPSFIGURE{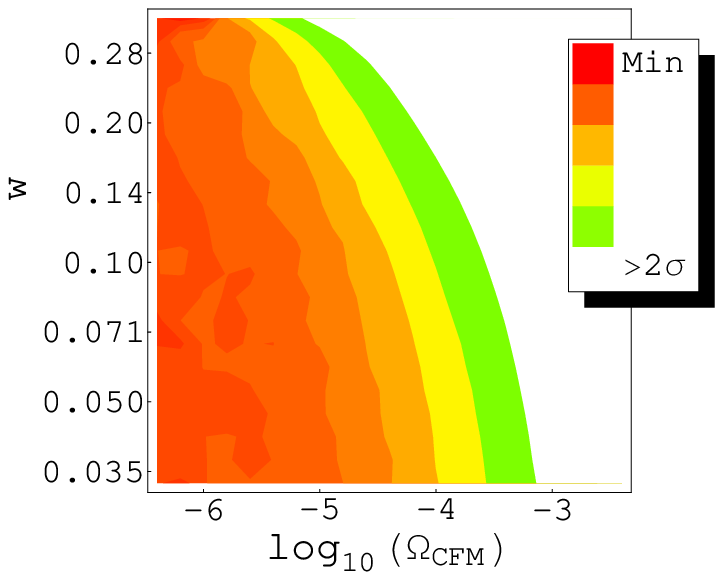,width=3in}{
The $\chi^2$ contours in the  $w$-$\Omega_{\rm WDM}$ plane.\label{chi2}
}

There are some obvious constraints our model must satisfy, namely that
the WDM component of WDF should not dominate the energy density of the
Universe at nucleosynthesis. 
Comparing the energy density of WDM to that of radiation, we find
\begin{equation}
\label{nucleoconstraint}
\frac{\rho_{\rm WDM}}{\rho_{\rm rad}} = \frac{\Omega_{\rm WDM}}{\Omega_{\rm rad}} (1+z)^{3w-1}.
\end{equation}
Since the CDM and WDE components can be neglected at nucleosynthesis, we find that we can obtain a relation between $w$ and $\Omega_{\rm WDM}$ by making use of \cite{kolbturner}
\begin{equation}
\label{nucleobound}
\rho_{\rm rad} = \frac{\pi^2}{30} g_{*}(T) T^4,
\end{equation}
where $g_{*}(T\sim 1\ {\rm MeV})=g_{*}^{\rm standard}+\Delta g_{*}$. The contribution of standard model particles in thermal equilibrium at $T\sim 1\ {\rm MeV}$ is $g_{*}^{\rm standard}=10.75$. Standard BBN light element abundances lead to the bound $g_{*}(T\sim 1\ {\rm MeV})\leq 12.50$, which in turn leads to:
\begin{equation}
\label{deltagstar}
\Delta g_{*} \equiv \frac{\Omega_{\rm WDM}}{\Omega_{\rm rad}} (1+z_{\rm Nuc})^{3w-1} g_{*}^{\rm standard}\leq 1.75\Rightarrow \frac{\Omega_{\rm WDM}}{\Omega_{\rm rad}} \leq 0.16 \ (2.36\times 10^{-10})^{3w-1}.
\end{equation}
Using the value $h=0.7$ for the Hubble parameter we find that getting the correct light element abundances imposes the following constraint: $\ln \Omega_{\rm WDM}\leq 11.0-66.5\ w$. 

As mentioned above, we considered three distinct experiments; 1) CMB
measurements from WMAP\cite{WMAP}, 2) matter power spectrum from the
2dF survey\cite{2df} and 3) type Ia supernovae
observations\cite{Riess}. We used CMB-Fast\cite{CMBFast} to compute the
scalar power spectrum and the transfer functions for our equation of
state. The transfer functions were then used to obtain the matter
power spectrum.

To fit to the supernovae data we integrated $d_L\!(z)\!=\!
(1\!+\!z)\int_0^z dz'/H\!(z')$ numerically to obtain the
red-shift/distance modulus relation for our model. We then
simultaneously fit all three experiments by constructing the combined
$\chi^2$. Although the full model has a number of parameters,
including many nuisance parameters, in practice the CMB data places
strong enough constraints on the domain of $\Omega_{\rm WDM}$ so that
we can fix most of the parameters to their best fit value from
$\Lambda$CDM, leaving us free to consider $w$ and $\Omega_{\rm
WDM}$. We calculated the combined $\chi^2$ on a grid of
$w$-$\Omega_{\rm WDM}$ values to identify the confidence regions and
found that the fit is primarily constrained by the CMB and the matter
\TABULAR{|l|c|c|r|l|}{
\hline
 & $w$ & $\Omega_{\rm WDM}$ & $\chi^2$ &\emph{Description} \\
\hline
\hline
$\Lambda$CDM & - & - & 427.291 & Concordance model \\
Set 1    & 0.316228 & $5.012 \times 10^{-7}$ & 427.263 & Best fit point (approximate) \\
Set 2    & 0.031623 & $1.995 \times 10^{-4}$ & 428.806 & Large $\Omega_{\rm WDM}$ good fit \\
Set 3    & 0.200000 & $1.000 \times 10^{-3}$ & 2425.883 & Sample point outside $2\sigma$ contour \\
Set 3* & 0.300000 & $2.850 \times 10^{-1}$ & $>$10000 & $\Omega_{\rm CDM}=0.004$ \\
\hline
}{\label{values}$(w,\Omega_{\rm WDM})$ points corresponding to figures
(\ref{snIa}),(\ref{cmb}) and (\ref{mps}).}

power spectrum, both of which generate similar confidence regions. The
supernova data places only rather weak constraints on our model, as it
only probes low $z$ where the model looks identical to $\Lambda$CDM.
\EPSFIGURE{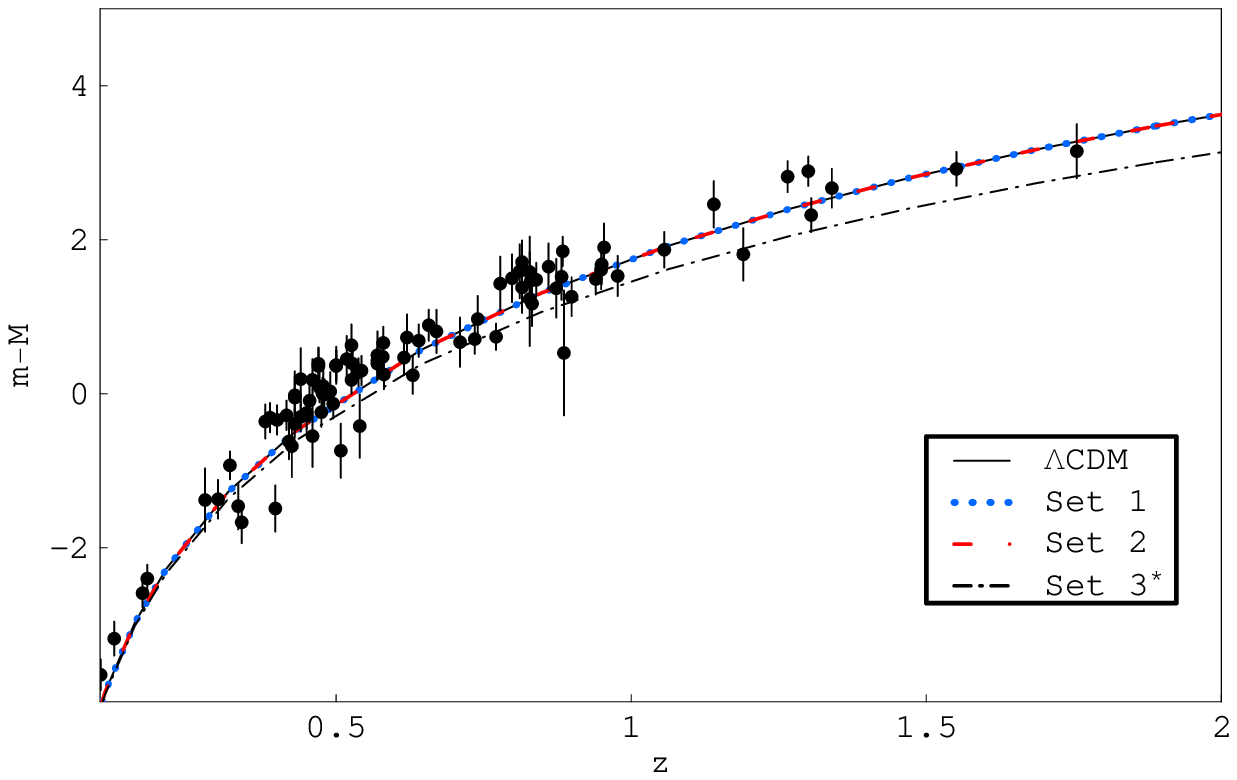,width=5in}{\label{snIa} The distance modulus for a
WDF dominated universe.  } Our best fit model has a combined $\chi^2$
statistically equivalent to the $\chi^2$ for $\Lambda$CDM. Figure
(\ref{chi2}) indicates the confidence regions in the $w$- $\Omega_{\rm
WDM}$ plane; figures (\ref{snIa}), (\ref{cmb}) and (\ref{mps}) depict,
for the particular values of $w$ and $\Omega_{\rm WDM}$ listed in
table (\ref{values}), the redshift-distance modulus relation and the
CMB power spectrum and matter power spectrum respectively. Unless
otherwise noted the models assume standard values for cosmological
paramters: $\Omega_{\rm WDE}+ \Omega_{\rm WDM}\!=\!0.73,\ \Omega_{\rm
CDM}\!=\!0.226$ and $\Omega_b\!=\!0.046$.
\EPSFIGURE{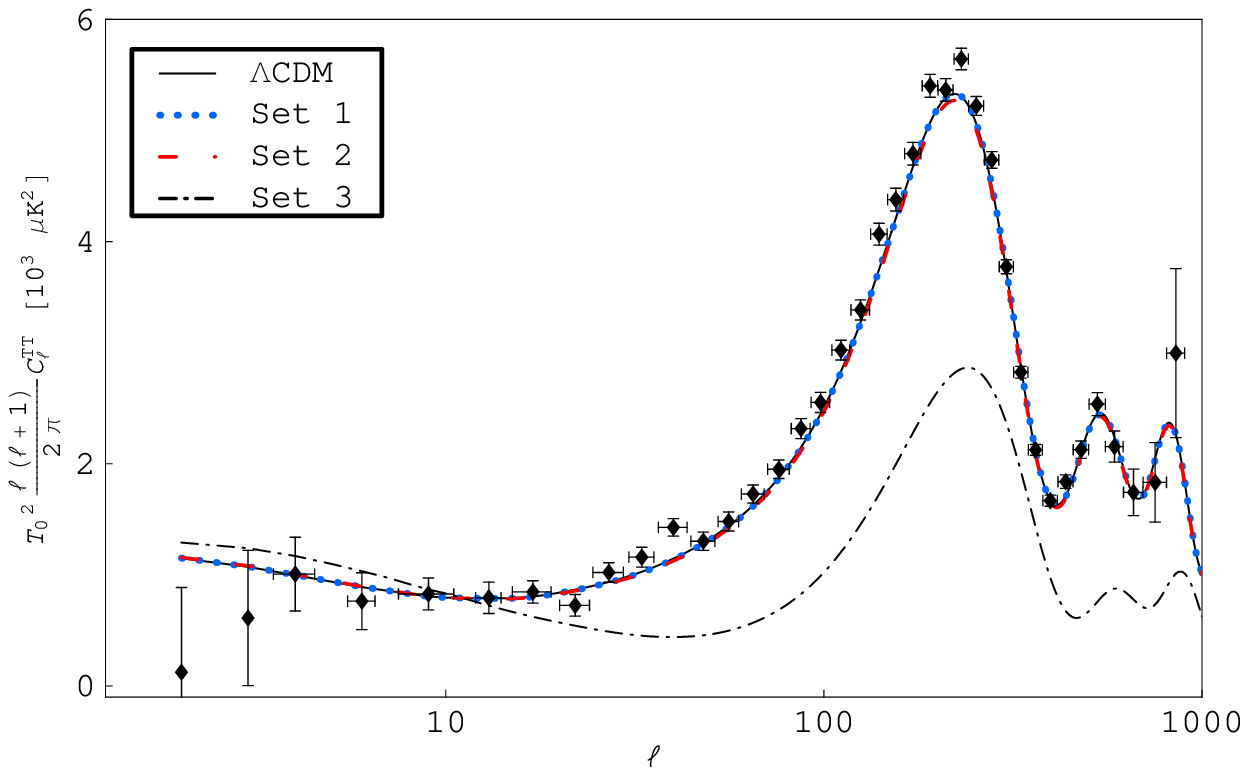,width=5in}{\label{cmb} 
The CMB temperature anisotropy as computed for a WDF dominated
universe.} 
From these figures we see that there are parameter values
for which {\em smooth} WDF provides as good a fit to all the available
data as the concordance model based on $\Lambda$CDM. The trend from
the $\chi^2$ contours is to drive $\Omega_{\rm WDM}$ to be relatively
small, of order $\Omega_{\rm WDM}\sim 10^{-6}-10^{-7}$, although
larger values can be used by paying a relatively small price in the
$\chi^2$ value.  While the scaling behavior of WDM is such that it
\EPSFIGURE{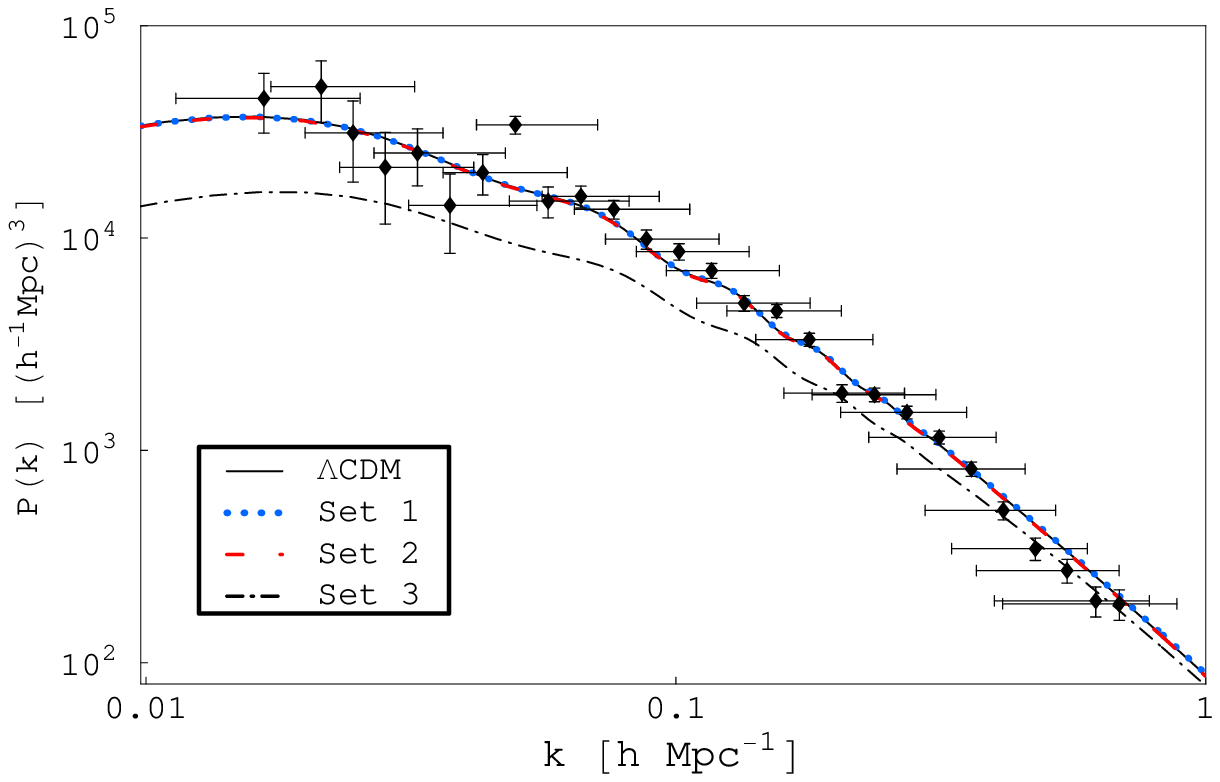,width=5in}{\label{mps} The matter power spectrum in
a WDF dominated universe.}  
would dominate over radiation at some point, the smallness of
$\Omega_{\rm WDM}$ means that this event lies sometime in the future
history of the Universe.

What has become increasingly clear from various parmeterizations of
the equation of state is that it must be very close to $\Lambda$CDM at
least up to $z\approx 1100$. This follows primarily from
the precision CMB and matter power spectrum measurements.  It is
important that the equation of state be physically reasonable and
approximate $\Lambda$CDM in the appropriate regime. However, we need
to be able to distinguish it from $\Lambda$CDM and hopefully this will
be possible by considering the growth of perturbations. This assumes
additional significance when we consider that the model differs
significantly from $\Lambda$CDM only behind the surface of last
scattering, a region quite inaccesible to current experiments.

Let's now turn to the possibility that the WDF is also affected by the primordial density perturbations generated in the early universe, perhaps through an inflationary phase. 
If we consider the hydrodynamics of a fluid governed by some general
equation of state characterized by $p=w(\rho)\rho$ and $v^2 \!\!=\!\!
dp/d\rho$ then linear perturbation theory tells us that the evolution of 
perturbations is governed by\cite{padmanabhan}
\begin{equation}
\label{perturbationeq}
\ddot{\delta}+H\dot{\delta}[2-3(2w-v^2)]
  -\frac{3}{2}H^2\delta [1-6v^2-3w^2+8w]
  =-\Bigl(\frac{kv}{a}\Bigr)^2\delta.
\end{equation}
where $\delta = \delta\rho/\rho$ and $k$ is the wavenumber of the
mode. In the $\Lambda$CDM case the cosmological constant is
homogeneous so that we need only consider the perturbations in the CDM component
\EPSFIGURE{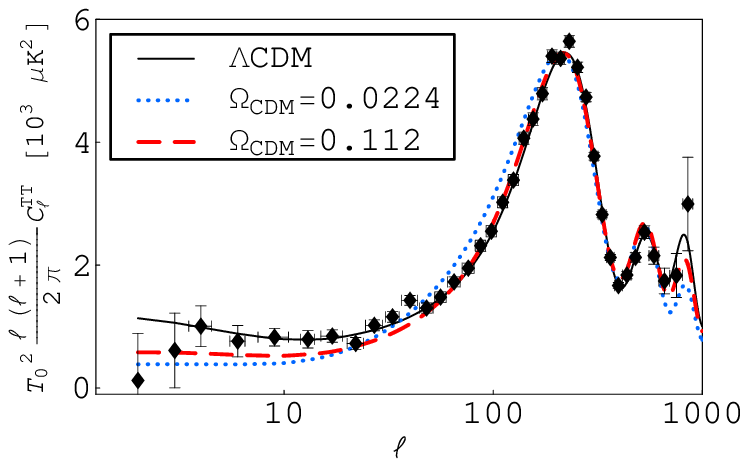,width=5in}{\label{pert-cmb}
CMB temperature anisotropy with perturbed WDF and CDM}
for which eq.(\ref{perturbationeq}) reduces to
\[
\ddot{\delta}+2H\dot{\delta}-\frac{3}{2}H^2\delta=0.
\]
If we allow perturbations in the WDF, the scenario is quite different
as the sound speed is a nonvanishing constant and $w(a)$ is no longer
constant. At late times when $w \rightarrow -1$ we find
\begin{equation}
\label{modeeq}
\ddot{\delta}+8H\dot{\delta}+15H^2\delta
  = -\Bigl(\frac{kv}{a}\Bigr)^2\delta .
\end{equation}
However, the fact that $w(a)$ {\em does} vary in time, will
significantly modify the behavior of some of the modes.  We recognize
from the RHS of Eq.~(\ref{modeeq}) a $k$ dependent suppression and that the
coefficients of the homogeneous equations differ substantially from
those of the $\Lambda$CDM case. We have used these equations in
CMB-Fast to calculate the CMB angular correlations, and the matter
power spectrum for various models.
\EPSFIGURE{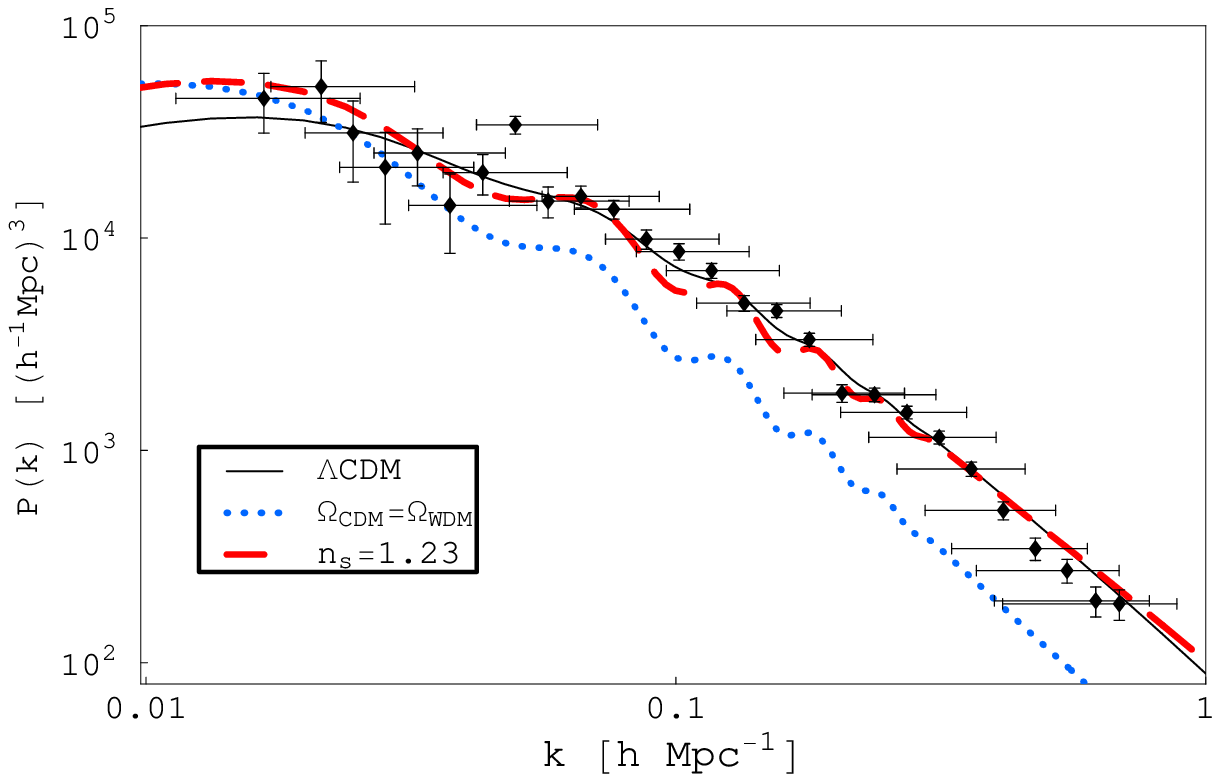,width=5in}{\label{pert-mps} Matter power
spectrum illustrating suppression. A tilted initial power spectrum can
compensate for the suppression.} 
The results are quite
interesting. If we normalize our calculated CMB power spectrum to the
location and height of the first peak we find that {\em even} for
$\Omega_{\rm WDM}\ge\Omega_{\rm CDM}$, our model agrees well with the
WMAP results as seen in figure (\ref{pert-cmb}). The change in $\chi^2$ from the concordance model is just
$\Delta\chi^2=5.3$ for $\Omega_{\rm WDM}=\Omega_{\rm CDM}$ . Furthermore, with
this normalization, our power spectrum also exhibits a low-$l$
suppression.

As one might suspect, the matter power spectrum provides more restrictions on the model,  as it
probes higher values of $k$. As shown in Fig.~(\ref{pert-mps}) we
observe an overall suppression int eh MPS, as well as an enhancement of the baryon
oscillations leading to $\Delta\chi^2=63.2$. The matter power spectrum
constrains $\Omega_{\rm CDM}$ to be greater that $\Omega_{\rm
WDM}$. However, this suppression can be compensated by increasing the
tilt of the power spectrum without significantly affecting the fit to
the WMAP data. This gives rise to a significant improvement of the fit, with
$\Delta\chi^2=3.7$.

We have considered both smooth and perturbed WDF fluids. However, it
would be very useful to have a field theoretic, microphysical model
that could give rise to the WDF equation of state. We could use this
to study the evolution of perturbations in a more detailed manner and
it's possible that entirely different properties may emerge when we
consider candidate microphysical theories.

Consider the following scalar field lagrangian \cite{ianetus}:
\begin{equation}
  \mathcal{L}\equiv\mathcal{L}(X)\quad\mathrm{where}\quad 
  X=g_{\mu\nu}\partial^{\mu}\phi\partial^{\nu}\phi.
\end{equation}
Such a Lagrangian can be made technically natural by the imposition of
the shift symmetry $\phi \rightarrow \phi+{\rm const.}$ Computing the
resulting stress-energy tensor we find
\begin{equation}
  {T^{\mu}}_{\nu}=-\mathcal{L}\;{\delta^{\mu}}_{\nu}
                 +2\partial^{\mu}\phi\partial_{\nu}\phi
                   \frac{d\mathcal{L}}{dX},
\end{equation}
from which we can obtain expressions for the energy density and pressure
once we make assumptions of spatial homogeneity and isotropy. These assumptions
imply that $\partial_i\phi=0$ and $X=\dot{\phi}^2$, so we can write
\begin{eqnarray}
  \rho & = & -\mathcal{L}+2X\frac{dL}{dX}, \nonumber \\
  p    & = & \mathcal{L}.
\end{eqnarray}
Now, assume that the ``vacuum'' state this theory finds itself in
has a non-zero value for $X$\cite{ghostcond}. Then we can arrive at a
choice of $\mathcal{L}$ that will give rise to the WDF equation of
state by having $\mathcal{L}$ satisfy the following differential
equation:
\begin{equation}
 \frac{2w}{1+w}X\frac{d\mathcal{L}}{dX}-\mathcal{L}-\frac{w}{1+w}\rho_{*}=0.
\end{equation}
If we define $\gamma=\frac{1+w}{w}$ and $M$ an arbitrary constant with
units of mass we can write the solution as:
\begin{equation}
\mathcal{L}(X)=(M^2)^{2-\gamma}X^{\frac{\gamma}{2}}-\frac{\rho_{*}}{\gamma}.
\end{equation}
The Euler-Lagrange equation following from this Lagrangian is
\begin{equation}
\ddot{\phi}+3w\frac{\dot{a}}{a}\dot{\phi}=0
\end{equation}
which implies $\dot{\phi}\propto a^{-3w}$. 

While this lagrangian does
serve our purpose, further work is required to investigate the nature
of perturbations about the homogenous solution. We begin this work by constructing the equation for field fluctuations around this background. This entails the 
replacing of $\phi$ with $\phi+\varphi$ and the evaluation of the contributions to
the stress energy and equation of motion, keeping only the first order terms in $\varphi$. We find
\begin{eqnarray}
\delta T^0_0&=&-(\gamma-1)\frac{\dot{\varphi}}{\dot{\phi}}(\rho+p),\\
\delta T^0_i&=&-\frac{k_i\varphi}{\dot{\phi}}(\rho+p)\quad\text{and}\\
\delta T^0_0&=&\frac{\dot{\varphi}}{\dot{\phi}}(\rho+p)\delta^i_j
\end{eqnarray}
for the first order contributions to the stress energy, where the
$(\rho+p)$ are the background homogeneous values. We need to choose a
gauge to write the equation of motion and here we pick the conformal
synchronous gauge\cite{mabert} so that we find
\begin{equation}
(\gamma-1)\ddot{\varphi}+(4-\gamma)\frac{\dot{a}}{a}\dot{\varphi}
+k^2\varphi+\frac{3}{2}\dot{\phi}\dot{h}=0,
\end{equation}
where $h$ represents the scalar metric perturbations. To go further we
have to encode these contributions in CMBFast and integrate them. We
will pursue this procedure in a future work.

To conclude, we've started with an equation of state that is extremely
simple and has a ``microphysical'' parameter $w$ that is {\em
positive} and bounded so that the adiabatic sound speed in this fluid
is less than $1$. Furthermore, this equation of state describes the
behavior of fluids as simple as water! From this we have extracted a
cosmological model that agrees with all available data and is as
statistically valid as the current favorite model, $\Lambda$CDM. We
would argue that this ushers in a new paradigm in understanding dark
energy in the sense that we can move away from demanding that the
microphysical $w$ be negative and we can certainly offer an
alternative to considering $w<-1$!  Furthermore, we have come up with
a proof of principle lagrangian that can describe this system.The next
step is to see how the perturbations from this class of theories
behaves, relative to $\Lambda$CDM, models, say. This work is in
progress.

Note: While we were writing this work up, \cite{russians} appeared
which also deals with this equation of state. Their viewpoint is
directed more towards the eventual fate of a Universe dominated by a
WDF-like equation of state and does not have much overlap with the
discussion in this paper.

\acknowledgments{
We thank R. B. Griffiths for comments about fluid equations of state and Ian Low for discussions about microphysical lagrangians. 
Our research was  supported in part by DOE grant DE-FG03-91-ER40682. 
}

\end{document}